# A Comparison of Named Entity Recognition Tools Applied to Biographical Texts

Samet Atdağ and Vincent Labatut

*Abstract—* Named entity recognition (NER) is a popular domain of natural language processing. For this reason, many tools exist to perform this task. Amongst other points, they differ in the processing method they rely upon, the entity types they can detect, the nature of the text they can handle, and their input/output formats. This makes it difficult for a user to select an appropriate NER tool for a specific situation. In this article, we try to answer this question in the context of biographic texts. For this matter, we first constitute a new corpus by annotating 247 Wikipedia articles. We then select 4 publicly available, well known and free for research NER tools for comparison: Stanford NER, Illinois NET, OpenCalais NER WS and Alias-i LingPipe. We apply them to our corpus, assess their performances and compare them. When considering overall performances, a clear hierarchy emerges: Stanford has the best results, followed by LingPipe, Illionois and OpenCalais. However, a more detailed evaluation performed relatively to entity types and article categories highlights the fact their performances are diversely influenced by those factors. This complementarity opens an interesting perspective regarding the combination of these individual tools in order to improve performance.

## I. Introduction

Identifying and categorizing strings of text into different classes is a process defined as *named entity recognition* (NER) [1]. Strictly speaking, a *named entity* is a group of consecutive words found in a sentence, and representing concepts such as persons, locations, organizations, objects, etc. For instance, in the sentence "Martin Schulz of Germany has been the president of the European Parliament since January 2012", "Martin Schulz", "Germany" and "European Parliament" are person, location and organization entities, respectively. Note there is no real consensus on the various types of entities. Although those are not exactly named entities, NER tools sometimes also handle numeric entities such as amounts of money, distances, or percentages, and hybrid entities such as dates (e.g. "January 2012" in the previous sentence).

Recognizing and extracting such data is a fundamental task and a core process of the natural language processing field (NLP), mainly for two reasons. First, NER is used directly in many applied research domains [1]. For instance, proteins and genes can be considered as named entities, and many works in medicine focus on the analysis of scientific articles to find out hidden relationships between them, and drive experimental research [2]. But NER is also used as a preprocessing step by more advanced NLP tools, such as relationship or information extraction [3]. As a result, a number of tools have been developed to perform NER.

NER tools differ in many ways. First, the methods they rely upon range from completely manually specified systems (e.g. grammar rules) to fully automatic machine-learning processes, not to mention hybrids approaches combining both. Second, they do not necessarily handle the same classes of entities. Third, some are generic and can be applied to any type of text [4-6], when others focus only on a specific domain such as biomedicine [7] or geography [8]. Fourth, some are implemented as libraries [6], and some take various other forms such as Web services [9]. Fifth, the data outputted by NER tools can take various forms, usually programmatic objects for libraries and text files for the others. There is no standard for files containing NER-processed text, so output files can vary a lot from one tool to the other. Sixth, tools reach different levels of performance. Moreover, their accuracy can vary depending on the considered type of entity, class of text, etc.

Because of all these differences, comparing existing NER tools in order to identify the more suitable to a specific application is a very difficult task. And it is made even harder by two other factors. First, most tools require the user to specify certain configuration settings, like choosing a dictionary. This leads to a large number of possible combinations, each one potentially corresponding to very different behaviors and performances. Second, in order to perform a reliable assessment, one needs an appropriate corpus of annotated texts. This directly depends on the nature of the application domain, and on the types of entities targeted by the user. It is not always possible to find such a dataset, and if none exist, then it must be designed manually, which is a long and difficult operation.

The work we present here constitutes a preliminary step in a larger research project, consisting in extracting spatiotemporal events from biographical texts. For this purpose, we need first to select an efficient NER tool. As mentioned earlier, there is no entity-annotated corpus for this usage, and the comparison of NER tools is difficult. To solve this problem, we constituted an appropriate corpus based on a selection of Wikipedia pages and developed a platform automating the comparison of NER tools. It includes a variant of classic performance measures we proposed to best fit our needs. Both the corpus and tool are publicly available under open licenses[1]. We then used our platform to compare the most popular free NER tools, and discuss their results.

The rest of this article is organized as follows. In the next section, we briefly review the selected NER tools. In section III we present the methods we used to evaluate their performance. We propose a new set of measures allowing to take partial matches into account. In section IV, we describe the corpus we created for this work and compare it to the

S. Atdag and V. Labatut are with the Department of Computer Science of the Galatasaray University, Istanbul, 34349, Turkey (e-mail: samet2@gmail.com, vlabatut@gsu.edu.tr).

---

[1] http://bit.gsu.edu.tr/compnet

existing ones. We then present and discuss, in section V, the performances obtained by the NER tools on these data. We conclude by highlighting the main points of our work, and discuss how it can be extended.

## II. EXISTING NER TOOLS

As mentioned before, many methods and tools were designed for named entity recognition. It is not possible to list them all here, but one can distinguish three main families [10]: hand-made rule-based methods, machine learning-based methods, and hybrid methods. The first use manually constructed finite state patterns [11]; the second treat NER as a classification process [10], and the third are a mix of those two approaches. We used three criteria for selecting appropriate NER tools. First, it must be publicly and freely available. Second, we favor proven tools, already well-established in the NER community. Last, due to our goal of finally identifying spatiotemporal events, we focused on tools able to handle at least Person, Location and Organization entities.

In the end, we selected four different tools. *Stanford Named Entity Recognizer* (SNER) [6] is based on linear chain conditional random fields. *Illinois Named Entity Tagger* (INET) [4] relies on several supervised learning methods: hidden Markov models, multilayered neural networks and other statistical methods. *Alias-i LingPipe* (LIPI) [12] uses $n$-gram character language models, trained through hidden Markov models and conditional random field methods. Several pre-trained models for the English language are provided with these three tools. Moreover, they all take the form of Java applications. The last selected tool differs in this point, since *OpenCalais* (OCWS) [9] is a Web service. Both LingPipe and OpenCalais are general tools, able to handle various other NLP tasks besides NRE. Moreover, both are commercial tools, but free licenses are available for academic use.

## III. EVALUATION METHODS

For a given text, the output of a NER tool is a list of entities and their associated types, and the ground truth takes the exact same form. In order to assess the tool performance, one basically wants to compare both lists. Different approaches can be used for this purpose, depending on the goal and context [1]. In this section, we first review the traditional approach, and then propose a variant adapted to our own context.

### A. Traditional Evaluation

The traditional evaluation relies on a set of counts classically used in classification: *True Positive* (TP), *False Positive* (FP) and *False Negative* (FN) counts. Those are used to process two distinct measures: Precision and Recall [13]. *Precision* is defined as the ratio $TP/(TP + FP)$. It corresponds to the proportion of detected entities which are correct. *Recall* is defined as $TP/(TP + FN)$. It is the proportion of real entities which were correctly detected. Both measures are complementary, in the sense they are related to type I (false alarm) and type II (miss) errors, respectively.

Comparing the estimated and actual lists of entities can be performed according to two distinct axes: *spatial* (position of the entities in the text) and *typical* (types of the entities). In terms of spatial performance, a TP is an actual entity whose position was correctly identified by the tool. A FP refers to an expression considered by the tool as an entity, but which does not appear as such in the ground truth. A FN is an actual entity the tool was not able to detect. Figure 1 presents an example of text extracted from Wikipedia and annotated. It contains 10 actual entities represented in boxes, and 9 estimated ones characterized by wavy underlines. In terms of exact matches, there are 5 TP (*Victor Charles Goldbloom*, *Montreal*, *Selwyn House*, *McGill University*, *New York*), 4 FP (*Canada*, *MD*, *Dr.Goldbloom*, *Medical Center*) and 5 FN (*Alton Goldbloom*, *Annie Ballon*, *Lower Canada College*, *Goldbloom*, *Columbia Presbyterian Medical Center*). This leads to a Precision of 0.56 and a Recall of 0.50.

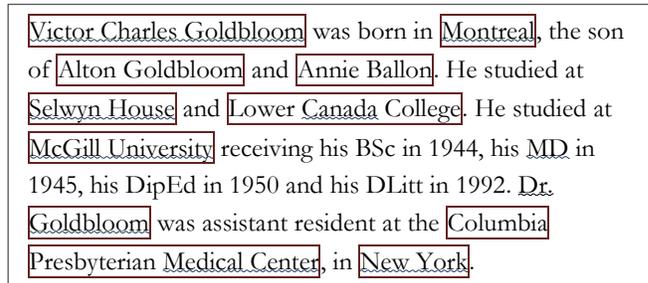

Figure 1. Example of annotated text

The interpretation of the counts is different when assessing the typical performance. TP correspond to entities whose type was correctly estimated. Due to the NER process, they consequently also correspond to entities whose position was identified at least partially correctly. FP are expressions considered by the tool as entities, but whose type was incorrectly selected, or which are not actual entities. FN are actual entities for which the tool selected the wrong type, or no type at all [1]. As an example, Table I contains the types of the entities from Figure 1. We count 7 TP (rows 1, 2, 5, 7, 9, 10 and 11 in Table I), 2 FP (rows 6 and 8) and 3 FN (rows 3, 4 and 6). Based on these counts, we get a Precision of 0.78 ands a Recall of 0.70.

TABLE I. TYPES OF THE ENTITIES IN FIGURE 1

| Reference entity | Reference | Estimation |
|---|---|---|
| Victor Charles Goldbloom | Person | Person |
| Montreal | Location | Location |
| Alton Goldbloom | Person | - |
| Annie Ballon | Person | - |
| Selwyn House | Organization | Organization |
| (Lower) Canada (College) | Organization | Location |
| McGill University | Organization | Organization |
| MD | - | Person |
| (Dr.) Goldbloom | Person | Person |
| (Col. Presb.) Medical Center | Organization | Organization |
| New York | Location | Location |

Recall and Precision can then be combined, for example using the F-Measure, in order to get a single score. Some authors even combine spatial and typical performances to get a single overall, somewhat easier to interpret, value. One of our goals with this work is to characterize the behavior of NER tools on biographical texts. To our opinion, combining the various aspects of the tool performance will result in a loss of very relevant information. To avoid this, we want to keep separated measures for space and types. For types, we decided to process Precision and Recall independently for each type. This allows assessing if the performance of a tool varies depending on the entity type. For instance, let us focus on Person entities from Table I. We count 2 TP (rows 1 and 9), 1 FP (row 8) and 2 FN (rows 3 and 4). For this specific type, we therefore get a Precision of 0.67 and a Recall of 0.50. For the spatial performance, we want to clearly distinguish partial and full matches. For this matter, in the next subsection we define variants of the traditional measures.

### B. Considering Partial Matches

The traditional approach used to assess spatial performance requires a *complete* match in order to count a TP: the boundaries of the estimated and actual entities must be exactly the same. However, in practice it is also possible to obtain *partial* matches [1], i.e. an estimated entity which intersects with an actual entity, but whose boundaries do not perfectly match. For example, in Figure 1 *Lower Canada College* is an actual entity, but the estimation only includes the word *Canada*. A partial match represents a significant piece of information: the NER tool detected something, even if it was not exactly the expected entity. Completely ignoring this fact seems a bit too strict to us. Moreover, in a later stage of our project, we will aim at developing a method to efficiently combine the findings of several NER tools, in order to improve the overall performance. From this perspective, it is important to consider the information represented by partial matches, and this is why we present an extension of the existing measures.

For this purpose, we propose alternative counts one can substitute to the previously presented ones. First, we need to count the *Partial Matches* (PM), i.e. the cases where the estimated entity contains only a part of the actual one. We consequently also need to consider the cases where the NER tool totally ignores the actual entity: we call this a *Complete Miss* (CM). The sum of PM and CM is equal to what was previously called FN. Another situation arises when the detected entities corresponds to no actual entity at all. We call this a *Wrong Hit* (WH). The sum of PM and WM is equal to FP. Finally, the last relevant case happens when we have a *Full Match* (FM). It exactly corresponds to a FP, but we decided to use a different name to define a consistent terminology. In the example from Figure 1, we have 5 FM (the entities previously considered as TP), 3 PM (*Lower Canada College*, *Dr.Goldbloom*, *Columbia Presbyterian Medical Center*), 1 WH (*MD*) and 2 CM (*Alton Goldbloom* and *Annie Ballon*).

We use our new counts to adapt the Precision and Recall measures. Regarding the numerator, we now have two different possibilities: FM or PM (instead of TP). For the Precision denominator, we need the total number of *estimated* entities, which amounts to $FM + PM + WH$ (and not $TP + FP$ anymore). For the Recall denominator, we use the total number of *actual* entities, which is $FM + PM + CM$ (and not $TP + FN$ anymore). We therefore obtain two kinds of Precision, which we coin *Full Precision* ($FM/(FM + PM + WH)$) and *Partial Precision* ($PM/(FM + PM + WH)$). Similarly, we have two kinds of Recall, called *Full Recall* ($FM/(FM + PM + CM)$) and *Partial Recall* ($PM/(FM + PM + CM)$). Additionally, a *Total* Precision (resp. Recall) can be obtained by summing Full and Partial Precisions (resp. Recalls). In the example of Figure 1, we get $Pre_F = 0.56$ and $Pre_P = 0.33$, so the Total Precision is 0.89. For the Recall, we have $Rec_F = 0.50$ and $Rec_P = 0.30$, resulting in a Total Recall of 0.80.

## IV. DESCRIPTION OF THE CORPUS

NER requires big amounts of data for both training and testing the tools. Most studies use some standard corpora generally designed for conferences or competitions, whereas some commercial tools are provided with their own data [12]. The *New York Times Annotated Corpus* [14] is a popular resource, constituted of manually annotated articles published in this journal. However, the access is conditional to the payment of a fee, and we decided to focus on freely available tools in this work. The *Message Understanding Conference* [15] proposed various corpora for NER. However, not all of them are freely available, and those which are focus on texts very different from biographies (terrorist reports, airplane crashes, etc.). The *National Institute of Standards and Technology* designed a NER corpus based on newswires [16], but it is not accessible from the web anymore. The *Conference on Computational Natural Language Learning* constituted NER corpora in 2002 and 2003, the latter in English [17]. However, all the articles are related to news, not biographies, and their access is commercial. In [18], four different corpora were constituted from emails, for the purpose of NER assessment. However, only Person entities were annotated. The *Automatic Content Extraction* corpora [19] are based on newswires, and their access requires to pay a fee.

TABLE II. NUMBER OF ENTITIES BY TYPE

| Person | Location | Organization | Date |
|---|---|---|---|
| 7330 | 2350 | 4611 | 4126 |

Due to the absence of a corpus meeting our needs and purpose, we designed a new one, specifically to assess NER tools on biographical texts. We first extracted more than 300 biographical articles from Wikipedia. We then cleaned and annotated 247 of them by hand. The corpus contains a total of 21364 annotated Person, Location, Organization and Date entities, as detailed in Table II.

TABLE III. NUMBER OF ARTICLES BY CATEGORY

| Politics | Science | Military | Art | Sports | Others |
|---|---|---|---|---|---|
| 94 | 48 | 11 | 34 | 25 | 37 |

The texts concern people from six categories of interest: Politics, Science, Military, Art, Sports, and other activities (medicine, law, etc.). The distribution of articles over categories is given in Table III. Note there are more politicians (from the 19[th] and 20[th] centuries) because the final goal of our spatiotemporal event extraction project primarily concerns this population.

## V. RESULTS AND DISCUSSIONS

We applied the NER tools described in section II on our corpus from section IV, using the measures presented in section III to assess their performance. The values obtained for the measures are displayed in Table IV and Table V, for spatial and typical evaluation, respectively. For NER tools proposing several pretrained models, we present only those having obtained the best performance. In order to study in details the behavior of the tested NER tools, we processed their performance not only for the whole corpus, but also by entity type and by article category.

*A. Overall Performance*

Let us first consider the overall performances. From a spatial perspective (Table IV), there is a clear hierarchy between the tools. When considering the total measures, i.e. the sum of full and partial measures, SNER comes second for Precision (0.88) and first for Recall (0.93). Moreover, the part of partial matches in these results is very low. LIPI has the third Precision (0.81) and the second Recall (0.89), but the part of partial matches is much higher.

INET is fourth for Precision (0.79) and third for Recall (0.78), and the share of partial matches are even more important (more than one third of the total performance). Note the fact the balance between full and partial matches changes from one tool to the other shows it is a relevant criterion for performance assessment. We manually examined the texts annotated by INET and found out this high level of partial matches has two main causes. First, many organization names include a location or a person name. INET tends to focus on them, rather than on the larger expression corresponding to the organization name. For example, in the expression *Toronto's Consulate General of the Netherlands*, INET detects the locations (*Toronto* and *Netherlands*). Second, INET has trouble detecting person names which include more than two words.

All previous three tools reach very comparable values for both measures. However, this is not the case for OCWS. This tool has the best Precision (0.91) but by far the worst Recall (0.61), with the smallest proportions of partial matches. The unbalance between the two measures means that OCWS is almost always right when detecting an entity, but also that it misses a lot of them.

With regards to the overall typical performances (Table V), the same hierarchy emerges between the tools. SNER has the second Precision (0.89) and the first Recall (0.92). It is followed by LIPI with the third Precision (0.82) and second Recall (0.88). INET reaches the fourth Precision (0.80) and third Recall (0.78). These values mean those tools perform relatively well, and are able to appropriately classify most entities. Moreover, their performances are balanced, which is not the case of OCWS. Exactly like for the spatial evaluation, we see OCWS reach the first Precision (0.91) but the last Recall (0.61). In words, on the one hand most of the entities it recognizes are correctly classified, but on the other hand it fails to correctly classify almost half the reference entities of the corpus.

*B. Performance by Entity Type*

Let us now comment the performances by entity type. For the spatial assessment, as shown in Table IV, SNER performs above its overall level when dealing with Person and Location entities (especially for the former). However, its performances are under it when it comes to Organizations: full match-based measures decrease, while partial match-based ones increase. The total measures stay relatively constant, though. An analysis of the annotated texts shows SNER has some difficulties in two cases, which mainly concern organizations. First, it tends to detect a full name followed by its abbreviation, such as in *Partido Liberación Nacional (PLN)*, as a single entity. Second, it sometimes splits names containing many words. For instance, in the phrase *Dr. Isaías Álvarez Alfaro*, it detects *Isaías* and *Álvarez Alfaro* as separate names. Finally, although it is less marked than for INET, SNER also sometimes mistakes person or location names in organization names. Regarding the typical performance, Person and Location entities are also slightly better handled: the former in terms of Precision and the latter in terms of Recall.

Concerning Person entities, the spatial performances of INET are very similar to the overall ones. For locations, the total precision decreases (due to less partial matches), whereas the recall increases (due to more full matches). In other words, INET is better as rejecting incorrect locations. For organizations, the total measures are similar to the overall level, but the share of partial matches is much higher. This means INET does not miss more Organization entities (compared to other types), but it has trouble precisely identifying their limits. In terms of typical performance, INET is clearly better on persons, both in terms of Precision and Recall. For locations, we can make the same observations than for the spatial performance, i.e. lower Precision and higher Recall compared to overall values.

For OCWS, compared to the overall results, we get similar values for locations, whereas those obtained for persons are slightly higher, and slightly lower for organizations. For all types, we observe the behavior already noticed at the overall level: Precision is high, comparable to the best other tools, whereas Recall is extremely low. A manual analysis of the annotated texts revealed OCWS has trouble handling acronyms, which mainly represent organizations in our corpus. In terms of typical performance,

TABLE IV. SPATIAL PERFORMANCE BY ENTITY TYPE AND ARTICLE CATEGORY

|  |  | SNER | | | | INET | | | | OCWS | | | | LIPI | | | |
|---|---|---|---|---|---|---|---|---|---|---|---|---|---|---|---|---|---|
|  |  | FP | PP | FR | PR | FP | PP | FR | PR | FP | PP | FR | PR | FP | PP | FR | PR |
| Overall |  | 0.78 | 0.10 | 0.83 | 0.10 | 0.53 | 0.26 | 0.52 | 0.26 | 0.81 | 0.10 | 0.55 | 0.06 | 0.64 | 0.17 | 0.70 | 0.19 |
| Type | Person | 0.87 | 0.05 | 0.89 | 0.05 | 0.56 | 0.28 | 0.54 | 0.27 | 0.87 | 0.07 | 0.56 | 0.04 | 0.79 | 0.11 | 0.81 | 0.12 |
|  | Location | 0.78 | 0.06 | 0.89 | 0.07 | 0.56 | 0.13 | 0.67 | 0.16 | 0.80 | 0.08 | 0.52 | 0.05 | 0.58 | 0.14 | 0.75 | 0.18 |
|  | Organization | 0.66 | 0.19 | 0.71 | 0.20 | 0.48 | 0.32 | 0.43 | 0.29 | 0.74 | 0.14 | 0.54 | 0.10 | 0.47 | 0.29 | 0.49 | 0.30 |
| Category | Art | 0.71 | 0.12 | 0.77 | 0.13 | 0.63 | 0.17 | 0.66 | 0.18 | 0.77 | 0.11 | 0.51 | 0.08 | 0.59 | 0.19 | 0.60 | 0.19 |
|  | Military | 0.75 | 0.15 | 0.80 | 0.16 | 0.50 | 0.24 | 0.47 | 0.23 | 0.75 | 0.16 | 0.44 | 0.09 | 0.64 | 0.20 | 0.58 | 0.18 |
|  | Politics | 0.77 | 0.12 | 0.80 | 0.12 | 0.47 | 0.33 | 0.45 | 0.32 | 0.80 | 0.10 | 0.45 | 0.06 | 0.66 | 0.18 | 0.63 | 0.17 |
|  | Science | 0.77 | 0.13 | 0.80 | 0.13 | 0.61 | 0.21 | 0.57 | 0.20 | 0.82 | 0.12 | 0.46 | 0.07 | 0.63 | 0.20 | 0.64 | 0.20 |
|  | Sports | 0.85 | 0.07 | 0.85 | 0.08 | 0.46 | 0.26 | 0.41 | 0.23 | 0.92 | 0.05 | 0.44 | 0.03 | 0.65 | 0.24 | 0.60 | 0.22 |
|  | Others | 0.73 | 0.15 | 0.76 | 0.15 | 0.57 | 0.26 | 0.55 | 0.25 | 0.80 | 0.12 | 0.49 | 0.07 | 0.56 | 0.25 | 0.56 | 0.25 |

Person entities are also more accurately classified, and the tool is slightly better at not misclassifying organizations.

The performance of LIPI is much better on persons than overall, for both Precision and Recall: this is true for both spatial and typical measures. For locations, we observe a decrease in Full Precision and an increase in Full Recall, also for both spatial and typical results. Our interpretation is that LIPI detects more incorrect locations, but misses less correct ones. For organizations, there is a clear decrease, in terms of both Precision and Recall, with a larger part of partial matches. This last observation can be explained by the fact LIPI tends to merge consecutive organizations.

*C. Performance by Article Category*

Certain article categories have an effect on the performance of certain tools. When considering SNER, there is no effect for the categories Military, Politics and Science. However, Art and Others lead to slightly lower performances, in terms of both space and types. On the contrary, the spatial performance is much higher than the overall level for Sports (and it is also true of the typical performance, at a lesser degree). This appears to be due to the fact the sport-related biographies generally contain a lot of person names, such as team-mates, opponent, coaches, etc. SNER is particularly good at recognizing person names, which is why its performances are higher for this category. Art-related articles contain many titles of artworks, which are generally confusing for NER tools: they often mistake them for organization names.

For INET, we observe a clear spatial performance increase for Art articles, which means it is not concerned by the previous observation. The performance is slightly better for Science and Sports, in the sense the proportion of full matches gets higher for both Precision and Recall (the total performance staying approximately equal). On the contrary, the values are lower for Military, Politics and Sports. One difficulty with military texts is the detection of army units (e.g. $2^{nd}$ *Stryker Cavalry Regiment*) as organizations. In terms of typical performance, the differences are strongly marked only for Art and Others, positively, and for Sport, negatively. So in Art articles, INET is better than usual, not only at identifying the limits of entities, but also at classifying them, whereas it is the opposite for Sport.

In terms of spatial performance, OCWS is not very sensitive to categories: the observed performances are very similar to the overall ones. The Sports category constitutes an exception though: total Precision stays the same, but the full Precision clearly increases, meaning OCWS is able to detect entities limits more accurately. This is certainly due to the presence of more person names, as already stated for SNER: OCWS gets its best performance on this entity type. The typical performances are more contrasted. The tool is clearly better on Science articles, for which its Recall is almost at the level of the other tools (0.63). On the contrary, the Recall is very low for Art, Others and especially Military (0.05). For the latter, it incorrectly classifies (or fail to detect) almost all the actual entities.

Like OCWS, the spatial performance of LIPI is not much affected by the article categories. For the Others category though, we observe a behavior opposite to that of OCWS for Sports: total Precision and Recall stay approximately constant, but the part of partial matches increases. It is difficult to interpret this observation, since this category corresponds to heterogeneous article themes. For the typical categories, we observe small variations. The classification is slightly better on Sports and slightly worse on Art.

*D. General Comments*

Several interesting conclusions can be drawn from our results and observations. First, even if the overall performances seem to indicate SNER as the best tool, it is difficult to rank them when considering the detailed performances. This puts in relief the fact single measures might be insufficient to properly assess the quality of NER tools and compare them. The different aspects we considered all proved to be useful to characterize the tools in a relevant way: partial matches, entity types, article categories.

As a related point, it turns out NER tools are affected by these factors in different ways. This is also why they are difficult to rank: none of them is the best on every type and category. As a consequence, these tools can be considered as complementary. For instance, if we consider types, then SNER is the best at recognizing persons. OCWS can be trusted when it recognizes locations and organizations, however is prone to missing a lot of them. On the contrary, LIPI is very good at not missing locations, but also incorrectly detect a lot of them. The differences are not as obvious for article categories, but this information can still be useful, e.g. SNER is much reliable when processing Sports articles. A set of voting rules could be manually derived to take advantage of these observations, or an automatic approach could be used, such as the training of a standard classifier, in order to combine outputs of individual NER tools, and increase the performance of the overall NER

TABLE V. TYPICAL PERFORMANCE RESULTS BY ENTITY TYPE AND ARTICLE CATEGORY

|  |  | SNER | | INET | | OCWS | | LIPI | |
| --- | --- | --- | --- | --- | --- | --- | --- | --- | --- |
|  |  | Precision | Recall | Precision | Recall | Precision | Recall | Precision | Recall |
| Overall | | 0.88 | 0.93 | 0.80 | 0.78 | 0.91 | 0.61 | 0.82 | 0.88 |
| Type | Person | 0.93 | 0.95 | 0.84 | 0.82 | 0.94 | 0.68 | 0.91 | 0.94 |
| | Location | 0.85 | 0.97 | 0.71 | 0.85 | 0.88 | 0.60 | 0.73 | 0.94 |
| | Organization | 0.85 | 0.92 | 0.80 | 0.74 | 0.88 | 0.69 | 0.76 | 0.80 |
| Category | Art | 0.83 | 0.92 | 0.80 | 0.86 | 0.90 | 0.34 | 0.78 | 0.81 |
| | Military | 0.91 | 0.96 | 0.81 | 0.78 | 0.96 | 0.05 | 0.85 | 0.77 |
| | Politics | 0.89 | 0.92 | 0.80 | 0.77 | 0.91 | 0.53 | 0.84 | 0.81 |
| | Science | 0.90 | 0.93 | 0.82 | 0.80 | 0.94 | 0.63 | 0.83 | 0.85 |
| | Sports | 0.93 | 0.93 | 0.73 | 0.64 | 0.96 | 0.51 | 0.89 | 0.83 |
| | Others | 0.87 | 0.91 | 0.83 | 0.81 | 0.92 | 0.19 | 0.81 | 0.80 |

system.

## VI. Conclusion

In this article, we focus on the problem of selecting an appropriate named entity recognition (NER) tool for biographic texts. Many NER tools exist, most of them based on generic approaches able to handle any kind of text. So, their performances on these specific data need to be compared in order to make a choice. However, existing corpora are not constituted of biographies. For this reason, we designed our own one and applied a selection of publicly available NER tools on it: Stanford NER [6], Illinois NET [4], OpenCalais WS [9] and LingPipe [12]. In order to highlight the importance of partial matches, we evaluated their performance using custom measures allowing to take them into account. Our results show a clear hierarchy between the tested tools: first Stanford NER, then LingPipe, Illinois NET and finally OpenCalais. The latter obtains particularly low Recall scores. When studying the detail of these performances, it turns out they are not uniform over entity types and article categories. Moreover, clear differences exist between tools in this regard. A tool like OpenCalais, which performs apparently much lower than the others (on these data), is still of interest because it can be good on niches, and therefore complete an otherwise better performing tool such as Stanford NER.

Our contribution includes four points. The first one is the constitution of a biographic corpus. It is based on articles of the English version of Wikipedia. We manually annotated 247 texts to explicitly highlight Person, Organization, Location, and Date entities. The second point is the definition of performance measures allowing to take partial matches into account. For this purpose, we modified the Precision, Recall and F-Measure traditionally used in text mining. The third point is the implementation of a platform allowing to benchmark NER tools. It is general enough to be easily extensible to other NER tools, corpora and performance measures. Our corpus and platform are both freely available online. The last point concerns the application of this platform to the comparison of four popular and publicly available NER tools.

This work can be extended in several ways. First, the size of the corpus could be increased, in order to get more significant results. This would also allow using a part of the corpus for training, and therefore obtain classifiers possibly more adapted to process biographies than the general ones we used here. However, article annotation is a very difficult and time-costly task. Second, the benchmark could involve more NER tools, so that the results reflect more completely the possible choices of the end user. Finally, the comparison we conducted here showed individual NER tools perform diversely on *bibliographic* texts. Their results are influenced by factors such as entity types and article categories. Moreover, they are not affected in the same way: some are better at recognizing locations, others at organizations, etc. Those tools can therefore be considered as complementary. Combining their outputs by giving them more or less importance depending on these factors seems like a promising way of improving the global NER performance. This could be achieved by defining a set of voting rules based on the observations we made during this study, or by training a classifier.

## Acknowledgment

The first version of our platform was developed by Yasa Akbulut, and the first version of our corpus was constituted by Burcu Küpelioğlu. Both tasks were conducted in our research team.

## References

[1] D. Nadeau, "Semi-Supervised Named Entity Recognition: Learning to Recognize 100 Entity Types with Little Supervision," Ottawa-Carleton Institute for Computer Science, School of Information Technology and Engineering, University of Ottawa, 2007.

[2] L. Tanabe, N. Xie, L. H. Thom, W. Matten, and W. J. Wilbur, "GENETAG: a tagged corpus for gene/protein named entity recognition," *BMC Bioinformatics,* vol. 6, p. S3, 2005.

[3] N. Bach and S. Badaskar, "A Review of Relation Extraction," in *Literature review for Language and Statistics II*, 2007.

[4] L. Ratinov and D. Roth, "Design challenges and misconceptions in named entity recognition," in *13th Conference on Computational Natural Language Learning*, 2009, pp. 147-155.

[5] F. Landsbergen, "Named Entity Work in IMPACT," presented at the IMPACT Final Conference 2011, 2011.

[6] J. R. Finkel, T. Grenager, and C. Manning, "Incorporating Non-local Information into Information Extraction Systems by Gibbs Sampling," in *43rd Annual Meeting on ACL*, 2005, pp. 363-370.

[7] R. Leaman and G. Gonzalez, "BANNER: an executable survey of advances in biomedical named entity recognition," *Pacific Symposium on Biocomputing,* vol. 13, pp. 652-663, 2008.

[8] B. Martins, M. Chaves, and M. J. Silva, "Assigning geographical scopes to Web pages," *LNCS,* vol. 3408, pp. 564-567, 2005.

[9] Thomson Reuters. (2008). *Calais Web Service*. Available: http://www.opencalais.com/

[10] A. Mansouri, L. Suriani Affendey, and A. Mamat, "Named Entity Recognition Approaches," *International Journal of Computer Science and Network Security,* vol. 8, pp. 339-344, 2008.

[11] G. D. Zhou and J. Su, "Named entity recognition using an HMM-based chunk tagger," in *40th Annual Meeting on ACL*, 2001, pp. 473-480.

[12] Alias-i. (2008, February 22, 2013). *LingPipe 4.1.0*. Available: http://alias-i.com/lingpipe

[13] E. F. T. K. Sang, "Introduction to the CoNLL-2002 shared task: language-independent named entity recognition," in *6th Conference on Natural language Learning*, 2002, pp. 1-4.

[14] E. Sandhaus. (2008). *The New York Times Annotated Corpus*. Available: http://www.ldc.upenn.edu

[15] R. Grishman and B. Sundheim, "Message Understanding Conference-6: a brief history," in *16th conference on Computational linguistics*, 1996, pp. 466-471.

[16] G. Doddington, A. Mitchell, M. Przybocki, L. Ramshaw, S. Strassel, and R. Weischedel, "The Automatic Content Extraction (ACE) Program: Tasks, Data, and Evaluation," in *4th International Conference on Language Resources and Evaluation*, 2004.

[17] E. F. T. K. Sang and F. de Meulder, "Introduction to the CoNLL-2003 shared task: language-independent named entity recognition," in *7th conference on Natural Language Learning*, 2003.

[18] E. Minkov, R. C. Wang, and W. W. Cohen, "Extracting personal names from email: applying named entity recognition to informal text," in *HLT/EMNLP*, 2005, pp. 443-450.

[19] Linguistic Data Consortium. (2005). *Automatic Content Extraction*. Available: http://projects.ldc.upenn.edu/ace/data/